\documentclass[preprintnumbers,article,amsmath,amssymb,floatfix,10pt,prd,twocolumn,
superscriptaddress,nofootinbib]{revtex4-2}
\usepackage{bm}
\usepackage{amsfonts}
\usepackage{latexsym}
\usepackage[latin1]{inputenc}
\usepackage{graphicx}
\usepackage{amsmath}
\usepackage{palatino}
\usepackage{mathpazo}
\usepackage{textcomp}
\usepackage{float}
\usepackage{booktabs}
\usepackage{dcolumn}
\usepackage{ragged2e}
\usepackage{hyperref}
\hypersetup{colorlinks,citecolor=blue}
\hypersetup{colorlinks=true,linkcolor=blue,filecolor=magenta,    urlcolor=blue}
\usepackage{amsmath}
\usepackage{subfigure}
\usepackage[]{natbib}
\usepackage{xcolor}
\usepackage{orcidlink}
\usepackage{epsfig}
\usepackage{caption}
\usepackage{subcaption}
\usepackage{commath}
\def\jnl@style{\it}
\def\aaref@jnl#1{{\jnl@style#1}}

\def\aaref@jnl#1{{\jnl@style#1}}

\def\aj{\aaref@jnl{AJ}}                   
\def\apj{\aaref@jnl{ApJ}}                 
\def\apjl{\aaref@jnl{ApJ}}                
\def\apjs{\aaref@jnl{ApJS}}               
\def\apss{\aaref@jnl{Ap\&SS}}             
\def\aap{\aaref@jnl{A\&A}}                
\def\aapr{\aaref@jnl{A\&A~Rev.}}          
\def\aaps{\aaref@jnl{A\&AS}}              
\def\mnras{\aaref@jnl{Mon.~Not.~Roy.~Astron.~Soc.}}             
\def\prd{\aaref@jnl{Phys.~Rev.~D}}        
\def\prc{\aaref@jnl{Phys.~Rev.~C}}  
\def\prl{\aaref@jnl{Phys.~Rev.~Lett.}}    
\def\qjras{\aaref@jnl{QJRAS}}             
\def\skytel{\aaref@jnl{S\&T}}             
\def\ssr{\aaref@jnl{Space~Sci.~Rev.}}     
\def\zap{\aaref@jnl{ZAp}}                 
\def\nat{\aaref@jnl{Nature}}              
\def\aplett{\aaref@jnl{Astrophys.~Lett.}} 
\def\apspr{\aaref@jnl{Astrophys.~Space~Phys.~Res.}} 
\def\physrep{\aaref@jnl{Phys.~Rep.}}      
\def\physscr{\aaref@jnl{Phys.~Scr}}       
\def\commat{\aaref@jnl{Comm.~Math.~Phys.}}              
\def\science{\aaref@jnl{Science}}               
\def\cqg{\aaref@jnl{Classical Quant.~Grav.}}            
\def\jpcs{\aaref@jnl{JPCS}}                                     
\def\ijmpd{\aaref@jnl{Int.~J.~Mod.~Phys.~D}}                    
\def\grg{\aaref@jnl{Gen.~Relat.~Gravit.}}               
\def\rpp{\aaref@jnl{Rep.~Prog.~Phys.}}          
\def\npa{\aaref@jnl{Nucl.~Phys.~A}}        
\def\lrr{\aaref@jnl{Living Rev.~Rel.}}                   
\def\jcap{\aaref@jnl{J.~Cosmology Astropart.~Phys.}}    
\def\rmp{\aaref@jnl{Rev.~Mod.~Phys.}}   
\def\epjc{\aaref@jnl{Eur.~Phys.~J.~C}} 
\def\plb{\aaref@jnl{~Phy.~Lett.~B}} 
\def\mpla{\aaref@jnl{Mod.~Phy.~Lett.~A}} 
\def\arxiv{\aaref@jnl{arxiv.org}}

\allowdisplaybreaks[1]

\begin{document}
\color{black}       
\title{Cosmological constraints on time-varying cosmological terms: A study of FLRW universe models with $\Lambda(t)$CDM cosmology}

\author{M. Koussour\orcidlink{0000-0002-4188-0572}}
\email[Email: ]{pr.mouhssine@gmail.com}
\affiliation{Department of Physics, University of Hassan II Casablanca, Morocco.}

\author{N. Myrzakulov\orcidlink{0000-0001-8691-9939}}
\email[Email: ]{nmyrzakulov@gmail.com}
\affiliation{L. N. Gumilyov Eurasian National University, Astana 010008,
Kazakhstan.}

\author{J. Rayimbaev\orcidlink{0000-0001-9293-1838}}
\email[Email: ]{javlon@astrin.uz}
\affiliation{New Uzbekistan University, Mustaqillik Ave. 54, Tashkent 100007, Uzbekistan.}
\affiliation{University of Tashkent for Applied Sciences, Gavhar Str. 1, Tashkent 100149, Uzbekistan.}
\affiliation{National University of Uzbekistan, Tashkent 100174, Uzbekistan.}
%
\begin{abstract}
This paper explores models of the FLRW universe that incorporate a time-varying cosmological term $\Lambda(t)$. Specifically, we assume a power-law form for the cosmological term as a function of the scale factor: $\Lambda(t)=\Lambda_{0} a(t)^{-\alpha}$, where $\Lambda_{0}$ represents the present value of the cosmological term. Then, we derive an exact solution to Einstein's field equations within the framework of $\Lambda(t)$CDM cosmology and determine the best-fit values of the model parameters using the combined $H(z)$ + SNe Ia dataset and MCMC analysis. Moreover, the deceleration parameter demonstrates the accelerating behavior of the universe, highlighting the transition redshift $z_{tr}$, at which the expansion shifts from deceleration to acceleration, with confidence levels of $1-\sigma$ and $2-\sigma$. In addition, we analyze the behavior of the Hubble parameter, jerk parameter, and $Om(z)$ diagnostic. Our analysis leads us to the conclusion that the $\Lambda(t)$CDM model is consistent with present-day observations.

\end{abstract}

\maketitle

\tableofcontents

\section{Introduction}\label{sec1}

Recent research has been dedicated to uncovering the origins of the accelerated expansion observed in the present universe. This accelerated expansion is supported by observational evidence, including astrophysical observations of Type Ia supernovae (SNe Ia) \cite{Riess/1998,Perlmutter/1999}, cosmic microwave radiation \cite{Komatsu/2011,Planck/2014}, baryonic acoustic oscillations \cite{Eisenstein2005,Percival/2007}, and measurements of the Hubble constant \cite{Farooq/2017,Yu_2018}. These observations collectively provide strong evidence for the phenomenon of cosmic acceleration. The accelerating expansion of the universe is attributed to a new form of energy with negative pressure, commonly referred to as dark energy (DE) \cite{Peebles}. DE constitutes a significant component of the universe's energy field \cite{Planck}. The role of DE in driving the accelerated expansion of the universe has been the subject of extensive research in astrophysics. However, the nature of DE remains a challenging problem in theoretical physics, presenting a significant puzzle for cosmologists and physicists alike. The cosmological constant, characterized by a time-independent equation of state $\omega=-1$, stands as the earliest and most straightforward candidate for DE. However, from a theoretical perspective, it encounters two significant challenges: the fine-tuning problem and the cosmic coincidence problem \cite{Weinberg/1989,Steinhardt/1999}. These issues have spurred further exploration into alternative models of DE that can offer more satisfactory explanations for the observed accelerated expansion of the universe. In addition to the cosmological constant, several other dynamical DE models have been proposed with time-dependent equations of state to explain the accelerated expansion of the universe. These models include quintessence \cite{RP}, phantom energy \cite{M.S.,M.S.-2}, K-essence \cite{T.C.,C.A.}, Chaplygin gas \cite{M.C.,A.Y.}, and tachyon  fields \cite{T.P.}. In addition, there are interacting DE models, such as holographic models and agegraphic models \cite{F2,S10,M2}, which offer alternative approaches to understanding the nature of DE and its role in the evolution of the universe. On the other hand, there are modified gravity models that have attracted many researchers recently \cite{H.A.,H.K.,Odintsov1,Odintsov2,T1,T2,Q0,Q1}. 

In the same context, a variety of models have been proposed in cosmology, incorporating time-varying $\Lambda(t)$ or vacuum decay, often with assumed time dependencies for $\Lambda(t)$ \cite{Oztas,Vishwakarma,Overduin1}.  These models explore different phenomenological decay laws for $\Lambda(t)$, obtained through geometric analysis \cite{Azri1,Azri2} or quantum mechanical reasoning \cite{Szydlowski}. Also, researchers have explored interactions between vacuum and matter using different approaches, comparing to recent cosmological data \cite{Bruni,Papagiannopoulos,Benetti1,Benetti2}. One promising approach to address the limitations of the $\Lambda(t)$CDM model is the RVM (Running Vacuum Model. This model is derived from quantum field theory in curved spacetime, expressing the vacuum energy density as a series of powers of the Hubble function $H$ and its derivatives for cosmic time $t$. The RVM has shown potential in fitting with cosmological observables, outperforming the $\Lambda(t)$CDM model in certain scenarios \cite{Sola2017, Peracaula2018, Tsiapi2019, Mavromatos2021, Peracaula2023, Peracaula2021, Peracaula2018b}. For further details on the RVM and recent theoretical developments, readers can refer to relevant literature \cite{Pulido2020,Pulido2022a,Pulido2022b,Pulido2023,Peracaula2022}. Overduin and Cooperstock \cite{Overduin} investigated the change in the scale factor in relation to a variable cosmological term, represented by forms like $\Lambda = A t^{-l}$, $\Lambda = B a^{-m}$, $\Lambda = C t^{n}$, and $\Lambda = D q^{r}$ (where $A$, $B$, $C$, $D$, $l$, $m$, $n$, $r$ are constants). Rezaei et al. \cite{Rezaei} employed phenomenological reasoning to parameterize the time-dependent behavior of $\Lambda(t)$, expressing it as a power series expansion of the Hubble parameter and its time derivatives: $\Lambda(t) \propto H, \Lambda(t) \propto \dot{H}, \Lambda(t) \propto H^2$. Following this, we consider models of the FLRW universe that include a time-varying cosmological term $\Lambda(t)$. We specifically adopt a power-law form for the cosmological term in terms of the scale factor: $\Lambda(t)=\Lambda_{0} a(t)^{-\alpha}$, where $\Lambda_{0}$ denotes the present value of the cosmological term. Recently, various noteworthy findings within the realm of $\Lambda(t)$CDM models using the power law with different values of $\alpha$ have appeared in references \cite{Lopez,Abdel,Chen,Calvao,Freese,Gasperini,Overduin2}. 

The present manuscript is organized as follows. 
In Section \ref{sec2}, we present the flat FLRW universe in the context of $\Lambda(t)$CDM cosmology. In Section \ref{sec3}, we obtain an exact solution to Einstein's field equations within the framework of $\Lambda(t)$CDM cosmology. This solution is expressed in terms of the Hubble parameter, assuming a power-law form of $\Lambda(t)$. In Section \ref{sec4}, we analyze observational data to determine the best-fit values of the model parameters using a dataset that includes 31 points from the Hubble $H(z)$ dataset and 1048 samples from the SNe Ia dataset. Furthermore, we investigate the behavior of various cosmological parameters, including the Hubble parameter, the deceleration parameter, and the jerk parameter. 
In Section \ref{sec5}, we examine the behavior of the $Om(z)$ diagnostic, which helps differentiate between various DE models based on the values constrained by observational data. Finally, we discuss our findings in Section \ref{sec6}.

\section{Overview of $\Lambda(t)$CDM cosmology}\label{sec2}

Let's start with the assumption that the universe can be described by the homogeneous (meaning the universe is uniform in its distribution of matter), isotropic (meaning it looks the same in all directions), and spatially flat FLRW metric, expressed as \cite{Ryden}
\begin{equation} \label{FLRW}
ds^2 = -dt^2 + a(t)^2 \left[ dr^2 + r^2(d\theta^2 + \sin^2\theta d\phi^2) \right], 
\end{equation}
where $t$ is the cosmic time, $r$ is the comoving radial distance, $\theta$ is the polar angle, $\phi$ is the azimuthal angle, and $a(t)$ is the scale factor of the Universe which depends on time.  This metric is a cornerstone of modern cosmology and is a solution to Einstein's field equations in GR under the assumptions of homogeneity and isotropy on large scales. The scale factor $a(t)$ encodes the expansion of the universe, with $a(t_0)=1$ representing the present scale of the universe at $t=t_0$. The dynamics of the Universe, including its expansion history and the evolution of its contents (such as matter, radiation, and DE), are governed by the behavior of the scale factor $a(t)$ as a function of cosmic time.

In addition, the matter content of the Universe is assumed to consist of a perfect fluid, for which the energy-momentum tensor is given by
\begin{equation}
\mathcal{T}_{\mu \nu }=(\rho +p)u_{\mu }u_{\nu }+pg_{\mu \nu }. \label{EMT}
\end{equation}%

Here, $\rho$ represents the energy density of the fluid, $p$ stands for the fluid's pressure, $u^{\mu}=(1,0,0,0)$ denotes the 4-velocity of the fluid elements, and $g_{\mu \nu}$ is the metric tensor. In this study, we concentrate on the late-time universe, where we neglect the impact of radiation, represented by $\rho_{r}=0$. Consequently, the total density $\rho$ and total pressure $p$ are described as $\rho=\rho_{m}+\rho_{\Lambda}$ and $p=p_m+p_{\Lambda}$, respectively.  Here, $ \rho_m$ represents the density of non-relativistic matter, $\rho_{\Lambda}$ represents the density of DE (often associated with the cosmological term, which can vary with time  $\Lambda(t)$), $p_m=0$ is the pressure associated with non-relativistic matter, and $p_{\Lambda}$ is the pressure associated with DE. 

Initially, the cosmological term $\Lambda$ was often considered a constant of nature. This perspective is rooted in the Einstein field equations, which can be written as
\begin{equation}
G_{\mu\nu} + \Lambda \, g_{\mu\nu} = \kappa T_{\mu\nu} ,
\label{EFEs}
\end{equation}
where $\kappa=8 \pi G=1$ and $G_{\mu \nu}=R_{\mu \nu}- \frac{1}{2} R g_{\mu \nu}$ is the Einstein tensor. By calculating the covariant divergence of equation (\ref{EFEs}), bearing in mind that the vanishing covariant divergence of the Einstein tensor is ensured by the Bianchi identities, and presuming that the energy-momentum tensor adheres to the conservation law $\nabla^{\nu}T_{\mu\nu}=0$, we conclude that the covariant divergence of $\Lambda g_{\mu\nu}$ must also vanish, implying that $\Lambda$ is a constant. This reasoning firmly places $\Lambda$ on the left-hand side of the field equations, providing a geometrical interpretation of the cosmological term.

However, a more recent approach, exemplified by \cite{Overduin}, has been to treat $\Lambda$ as part of the matter content of the universe rather than a constant. This is achieved by moving $\Lambda$ to the right-hand side of the field equations:
\begin{equation}
G_{\mu\nu} = \kappa \tilde{T}_{\mu\nu},
\end{equation}
where $\tilde{T}_{\mu\nu} \equiv T_{\mu\nu} - \frac{\Lambda}{\kappa} \, g_{\mu\nu}$ is the effective energy-momentum tensor. This reinterpretation allows for the possibility that $\Lambda$ may vary, as long as the effective energy-momentum tensor $\tilde{T}_{\mu\nu}$ satisfies energy conservation \cite{Overduin}.

Now, by using the metric and the Einstein field equations, we derive the standard Friedmann equations as \cite{Myrzakulov,Macedo},
\begin{eqnarray}
\label{F1}
3H^{2}&=&\rho_{m}+\rho_{\Lambda}, \\
2{\dot{H}}+3H^{2}&=&-p_{\Lambda}, 
   \label{F2} 
\end{eqnarray}
where $H=\frac{\dot{a}}{a}$ is the Hubble parameter, which represents the current rate of expansion of the universe.

In the present manuscript, we assume the EoS of the vacuum to be $\omega_{\Lambda}=-1$. This choice implies that $p_{\Lambda}=-\rho_{\Lambda}=-\Lambda(t)$ \cite{Macedo}. However, a recent finding regarding the RVM is that the EoS of vacuum evolves with cosmic history \cite{Moreno}. With the given assumptions, we can express the second Friedmann equation (\ref{F2}) as \cite{Socorro, SolaPeracaula2023Arxiv}
\begin{equation} \label{F22}
2{\dot{H}}+3H^{2}=\Lambda(t).
\end{equation}

Also, we can express the continuity equation for the matter and vacuum energy components of the universe as follows:
\begin{align}
\dot{\rho}_{m} + 3H{\rho_{m}} &= Q,\label{rhom}\\
\dot{\rho}_{\Lambda} &= -Q.
\end{align}
where $Q$ represents the interaction term between pressureless matter and vacuum. Now, the dimensionless density parameters for the cosmological term and matter are defined as
\begin{equation}\label{3j}
  \Omega_{\Lambda 0}=\frac{\Lambda_0}{3H_{0}^2}, \ \  \   \Omega_{m0}=(1-\Omega_{\Lambda 0}).
\end{equation}

\section{Time-varying cosmological term model} \label{sec3}

To solve Eq. (\ref{F22}) for the Hubble parameter $H$, we require an additional equation. Therefore, we assume a well-motivated power-law form for the cosmological term as a function of the scale factor \cite{Overduin}, 
\begin{equation}
\label{La}
    \Lambda(t)=\Lambda_{0} a(t)^{-\alpha},
\end{equation}
where $\Lambda_{0}$ represents the present value of the cosmological term. The case where $\alpha$ has received significant attention in the literature and is motivated by dimensional arguments \cite{Lopez,Abdel,Chen,Calvao}. Another group of researchers has focused on values of $m$ approximately equal to $4$, where the cosmological term behaves similarly to ordinary radiation \cite{Freese,Gasperini,Overduin}.

Furthermore, by using the relation $1+z=\frac{1}{a(t)}$, we can establish the connection between the cosmic time $t$ and the redshift $z$ as described below,
\begin{equation}
\label{dt}
    \frac{d}{dt}=-H(z)(1+z) \frac{d}{dz}.
\end{equation}

Therefore, the time derivative of the Hubble parameter can be expressed in the form of
\begin{equation}
\label{dt}
\dot H=-H(z)(1+z) \frac{dH(z)}{dz}.
\end{equation}

Using Eqs. (\ref{F22}), (\ref{La}), and (\ref{dt}), we derive the expression for the Hubble parameter $H(z)$ in terms of the redshift $z$,
\begin{equation}
    H(z)=H_0 \sqrt{\frac{(\alpha -3 \Omega_{m0})(1+z)^3}{\alpha -3}+\frac{3 (\Omega_{m0}-1) (1+z)^{\alpha }}{\alpha -3}},
\end{equation}
where $H_0$ represents the present-day Hubble value, corresponding to $z=0$. In particular, when $\alpha=0$, the expression for $H(z)$  simplifies to 
\begin{equation}
    H(z)=H_0 \sqrt{\Omega_{m0}(1+z)^3+(1-\Omega_{m0})},
\end{equation}
which corresponds to a standard $\Lambda$CDM model. Hence, the model parameter $\alpha=0$ serves as a useful indicator of how much the current model deviates from the $\Lambda$CDM model.

To describe whether the cosmological expansion is accelerating or decelerating, we introduce the deceleration parameter $q$, which is defined as
\begin{equation} \label{q}
q=-1-\frac{\dot{H}}{H^2}= -1+\frac{(1+z)}{H(z)}\frac{dH(z)}{dz}.
\end{equation}

In the present model, the deceleration parameter is formulated as
\begin{equation}
    q(z)=-1+\frac{3 (\alpha -3 \Omega_{m0})(1+z)^3+3 \alpha  (\Omega_{m0}-1) (1+z)^{\alpha }}{2(\alpha -3 \Omega_{m0}) (1+z)^3+6 (\Omega_{m0}-1) (1+z)^{\alpha }}.
\end{equation}

When the condition $q<0$ is met, the universe will transition into an accelerated phase of expansion. In addition, the jerk parameter is a higher-order cosmological parameter that characterizes the rate of change of the deceleration parameter with respect to time \cite{Sahni:2002fz}. It is defined as \cite{Visser:2004bf}
\begin{equation}
    j(z)=(2 q+1) q+(1+z) \frac{d q}{d z}.
\end{equation}

The jerk parameter provides information about the rate of change in the acceleration or deceleration of the cosmic expansion, offering insights beyond what is captured by the deceleration parameter alone. In various cosmological models, the transition from the decelerating to the accelerating phase of the universe occurs when the jerk parameter takes on a positive value ($j>0$). For instance, in $\Lambda$CDM model, the jerk parameter $j$ remains constant at $j=1$ \cite{Visser:2004bf}.

\section{Observational data} \label{sec4}

\subsection{Data and methodology}

Now, we evaluate the consistency of the $\Lambda(t)$CDM model with recent observational data, focusing on the observational Hubble $H(z)$ data and Type Ia Supernovae (SNe Ia).
\begin{itemize}
\item \textbf{$H(z)$ dataset}: 
In our analysis, we consider 31 Hubble $H(z)$ data points obtained using the differential age (DA) approach, which provides valuable insights into the expansion history of the universe \cite{Yu_2018, Moresco/2015, Sharov/2018}.

\item \textbf{SNe Ia dataset}: In addition, our analysis includes 1048 Type Ia supernovae (SNe Ia) luminosity distance estimates derived from various sources, such as the Pan-STARRS 1 (PS1) Medium Deep Survey, the Low-z, SDSS, SNLS, and HST missions, which are part of the Pantheon sample. These data provide a comprehensive view of the universe's expansion history and contribute to our assessment of the $\Lambda(t)$CDM model's validity \cite{Scolnic/2018, Chang/2019}.
\end{itemize}

In our analysis of cosmological observational data, we use the Markov Chain Monte Carlo (MCMC) sampling technique. Our approach builds upon previous studies by incorporating a broad range of data sources and applying more stringent priors on the model parameters. Specifically, we focus on the parameter space $\theta_{s}=(H_{0},\Omega_{m0},\alpha)$ and employ the \textit{emcee} library \cite{Mackey/2013} for parallelized MCMC sampling. Our analysis employs 100 walkers and 1000 steps to ensure robust and reliable results. By combining information from both the Hubble $H(z)$ data and the Pantheon sample of SNe Ia, we extract valuable insights into the model parameters and their associated uncertainties, enhancing our understanding of the universe's expansion history.

We define the $\chi^2$ function for the joint $H(z)$+SNe Ia dataset as a measure of the goodness-of-fit for our model,
\begin{equation}
\chi^{2}_{joint}=\chi^{2}_{H(z)}+\chi^{2}_{SNe Ia},
\end{equation}
where
\begin{equation}
\chi^{2}_{H(z)} = \sum_{i=1}^{31} \frac{\left[H(\theta_{s}, z_{i})-
H_{obs}(z_{i})\right]^2}{\sigma(z_{i})^2},
\end{equation}
and
\begin{equation}
\chi^{2}_{SNe Ia} =\sum_{i,j=1} ^{1048} \Delta \mu_{i} \left(
C_{SNe Ia}^{-1}\right)_{ij} \Delta \mu_{j}.
\end{equation}

In this context, $H(z_{i})$ denotes the theoretical value of the Hubble parameter for a specific model at different redshifts $z_{i}$, $H_{obs}(z_{i})$ denotes the observed value of the Hubble parameter, $\sigma(z_{i})$ denotes the observational error, $\Delta \mu_{i}=\mu_{\rm th}-\mu_{\rm obs}$ denotes the difference between the theoretical and observed distance modulus, and $C_{SNe Ia}^{-1}$ is the inverse of the covariance matrix of the Pantheon sample, which characterizes the correlations between different measurements. For further details on these definitions and their application, please refer to the following references: \cite{Koussour1,Koussour2,Koussour3,Koussour4,Koussour5}.

\subsection{Result analysis and discussions}

\begin{figure}[h]
\centerline{\includegraphics[scale=0.6]{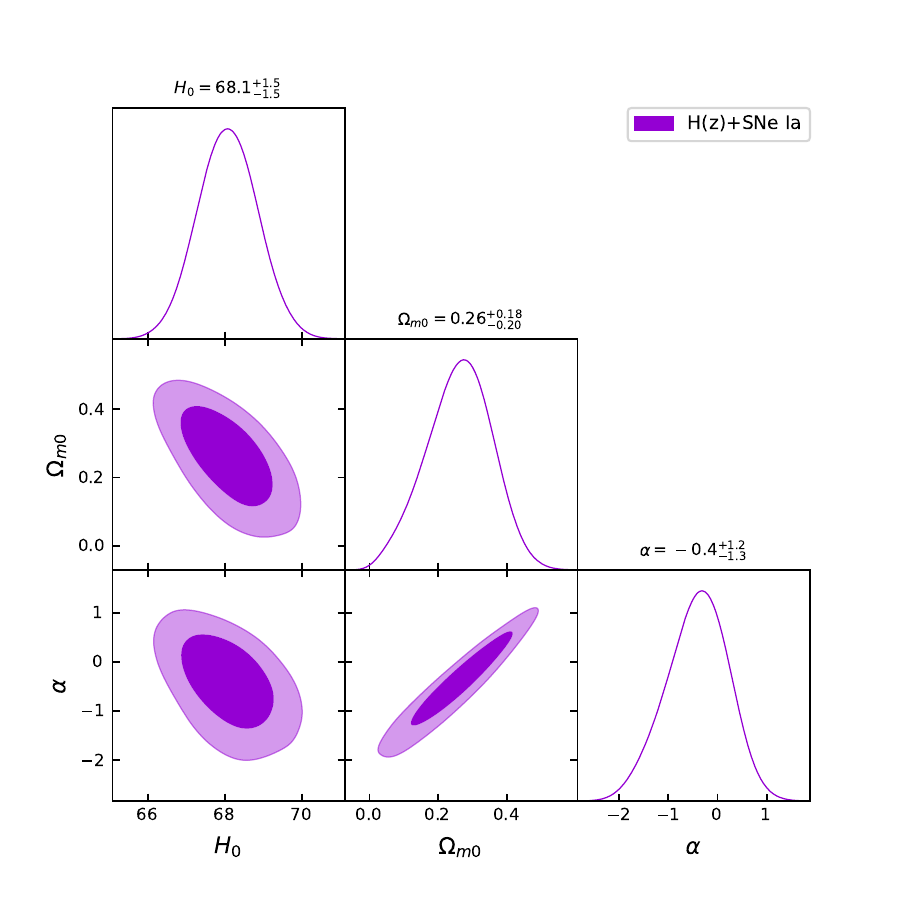}}
\caption{
The plot displays the best-fit values of the model parameters $\theta_{s}=(H_{0},\Omega_{m0},\alpha)$ obtained from a combined analysis of 31 points of $H(z)$ data and 1048 points of the SNe Ia dataset. The confidence levels of $1-\sigma$ and $2-\sigma$ are used to show the range of parameter values consistent with the observational data.}
\label{Combine}
\end{figure}

The model parameters are constrained by minimizing their respective $\chi^2$ values using MCMC with the \textit{emcee} library. The outcomes are summarized in Tab. \ref{tab}, which showcases the best-fit values and their associated uncertainties. Furthermore, Fig. \ref{Combine} displays the $1-\sigma$ and $2-\sigma$ contour plots for the combined observational data. These visual representations illustrate the relationships between parameters and identify the favored regions of parameter space based on the data. Here, we derive the Hubble constant as $H_0 = 68.1_{-1.5}^{+1.5}$ $km/s/Mpc$ by combining observational data from $H(z)$ and SNeIa. Recent studies have highlighted discrepancies, known as the Hubble tension, between the measurements of the Hubble parameter from the Planck satellite and other independent cosmological probes. The Planck collaboration \cite{Planck} estimated $H_0 = 67.4 \pm 0.5$ $km/s/Mpc$, while the HST team found $H_0 = 74.03 \pm 1.42$ $km/s/Mpc$ \cite{Riess:2019cxk}. This results in a significant tension of 4.4$\sigma$ between the two measurements. Further discussions on the Hubble tension and potential solutions are presented in Ref. \cite{DiValentino}. Comparing our results with those of the Planck collaboration \cite{Planck}, we find a tension of 0.44$\sigma$ in $H_0$ from Planck results \cite{Planck} based on our analysis of $H(z)$ and SNeIa. Additional investigations on the Hubble tension can be found in Refs. \cite{Yang:2021eud,DiValentino1}.

\begin{table}[h]
\begin{center}
\begin{tabular}{l c c c c}
\hline 
$dataset$              & $H_{0}$ ($km/s/Mpc$)& $\Omega_{m0}$ & $\alpha$ \\
\hline
$Priors$ & $(60,80)$  & $(0,1)$  & $(-10,10)$ \\

$H(z)+SNe Ia$   & $68.1_{-1.5}^{+1.5}$  & $0.26_{-0.20}^{+0.18}$  & $-0.4^{+1.2}_{-1.3}$ \\

\hline
\end{tabular}
\caption{An overview of the MCMC results obtained from the combined dataset.}
\label{tab}
\end{center}
\end{table}

\begin{figure}[h]
\centering
\includegraphics[scale=0.7]{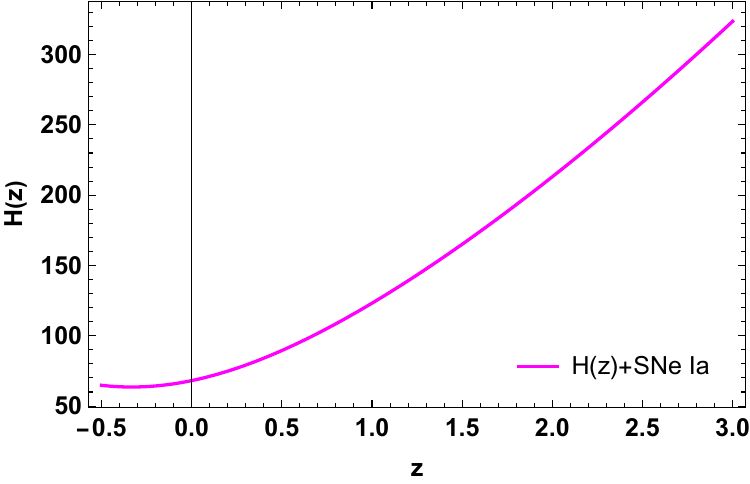}
\caption{Profile of the Hubble parameter based on the constraints obtained from the combined $H(z)$+SNe Ia dataset.}
\label{F_H}
\end{figure}

\begin{figure}[h]
\centering
\includegraphics[scale=0.7]{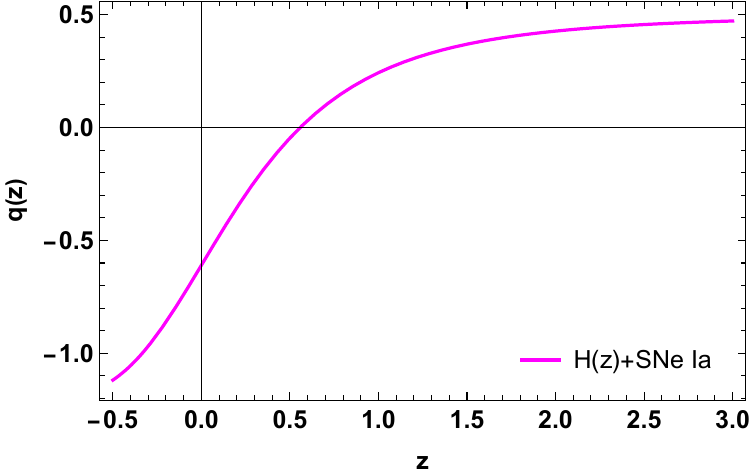}
\caption{Profile of the deceleration parameter based on the constraints obtained from the combined $H(z)$+SNe Ia dataset.}
\label{F_q}
\end{figure}

\begin{figure}[h]
\centering
\includegraphics[scale=0.7]{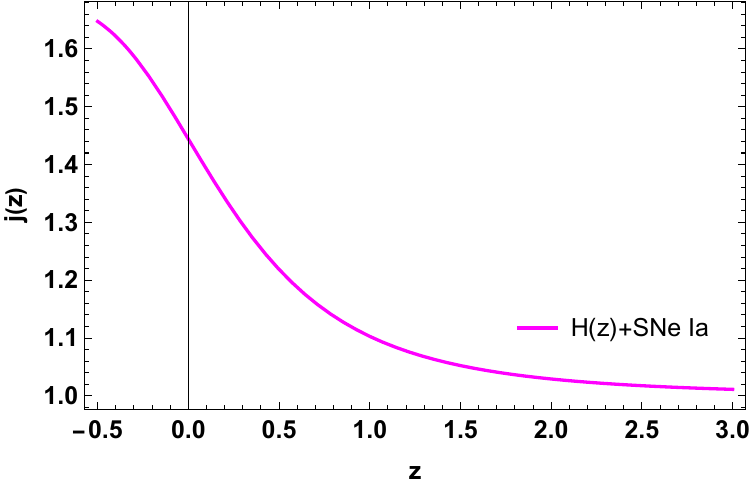}
\caption{Profile of the jerk parameter based on the constraints obtained from the combined $H(z)$+SNe Ia dataset.}
\label{F_j}
\end{figure}

The figures (Figs. \ref{F_H}, \ref{F_q}, and \ref{F_j}) depict the variations of the Hubble function $H(z)$, the deceleration parameter $q(z)$, and the jerk parameter $j(z)$ as functions of redshift $z$. These plots were generated using the values of the model parameters derived from the combined $H(z)$+SNe Ia dataset, providing information about the evolution of these cosmological quantities across different redshifts $(-0.5,3)$.

As depicted in Fig. \ref{F_H}, the Hubble function exhibits a monotonically increasing behavior with respect to redshift (or equivalently, a monotonically decreasing behavior with respect to time). The evolution of $H(z)$ is notably influenced by the model parameters. In particular, when $\alpha=0$, the model closely resembles the standard $\Lambda$CDM model in its behavior and characteristics.

The deceleration parameter $q(z)$, as illustrated in Fig. \ref{F_q}, exhibits a significant dependence on the model parameters. The plot in Fig. \ref{F_q} demonstrates a smooth transition of $q(z)$ from a decelerated phase ($q>0$) to an accelerated phase ($q<0$) of the universe's expansion at the transition redshift $z_{tr}=0.5^{+0.02}_{-0.02}$ for the best-fit model. However, it is important to note that there is an uninspired phase of the universe as $z$ approaches -1, where $q(z)$ diverges further from the de-Sitter phase for the combined dataset. These results are consistent with findings from several independent studies (see \cite{Magana,Mamon1,Mamon2}), which suggest that the universe underwent a phase transition from decelerating to accelerating expansion at redshifts approximately between $z_{tr}=0.5$ and $z_{tr}=1$.

The jerk parameter, as depicted in Fig. \ref{F_j}, is often utilized to differentiate between various DE models. In this particular case, it exhibits consistently positive behavior throughout the entire history of the universe. The observations suggest that, in this model, the present value of $j$ exceeds $1$ for the combined $H(z)$+SNe Ia dataset, specifically $j_0=1.4^{+0.02}_{-0.01}$ \cite{Mamon3}. Thus, based on the present model, it is apparent that the considered dynamic DE model is the most likely explanation for the ongoing acceleration and warrants further investigation. However, it is important to note that $j(z)$ is not well constrained in this scenario, as it may be associated with the emergence of abrupt future singularities \cite{Dabrowski,Pan}. This aspect introduces additional complexity and highlights the need for more comprehensive studies to fully understand the dynamics of the universe in this context.

\section{$Om(z)$ diagnostics} \label{sec5}

In the field of cosmology, a geometric approach is utilized, where the Hubble parameter serves as a critical test for the $\Lambda$CDM model. Moreover, the $Om(z)$ diagnostic is employed to effectively distinguish between different DE models and the $\Lambda$CDM model by observing the variation in the slope of $Om(z)$ \cite{Sahni}. Its simplicity arises from its dependence solely on the first-order derivative of the scale factor, which makes it particularly easy to apply and interpret in cosmological studies. In the case of a spatially flat universe, its expression is given by
\begin{equation}
Om\left( z\right) =\frac{\left( \frac{H\left( z\right) }{H_{0}}\right) ^{2}-1%
}{\left( 1+z\right) ^{3}-1},    
\end{equation}
where $H_0$ represents the present value of the Hubble parameter. A positive or negative slope of the $Om(z)$ diagnostic identifies a quintessence or phantom model, respectively. Moreover, a constant slope with respect to redshift indicates a DE model corresponding to the cosmological constant. 

\begin{figure}[h]
\centering
\includegraphics[scale=0.7]{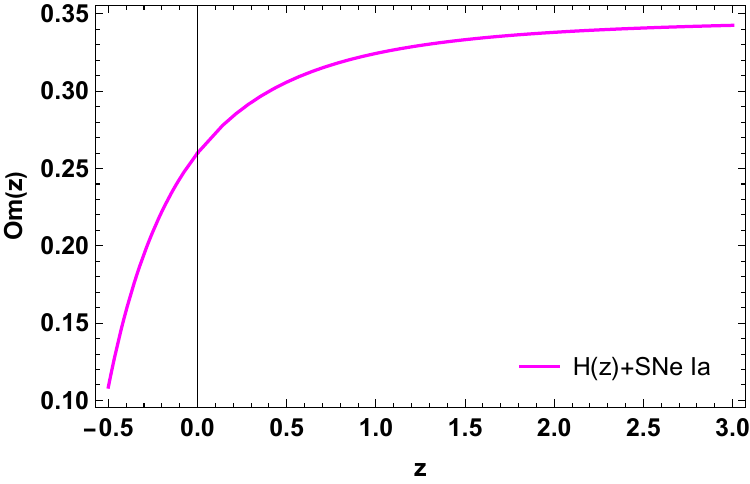}
\caption{Profile of the $Om(z)$ diagnostic based on the constraints obtained from the combined $H(z)$+SNe Ia dataset.}
\label{F_Om}
\end{figure}

Fig. \ref{F_Om} illustrates that the $Om(z)$ diagnostic exhibits a positive slope for the best-fit values of the model parameters constrained from the combined $H(z)$ + SNe Ia dataset. Consequently, based on the $Om(z)$ test, we can conclude that $\Lambda(t)$CDM model demonstrates phantom-type behavior, which refers to a scenario where the equation of state parameter for DE is less than -1.

\section{Conclusion} \label{sec6}

The confirmed accelerated expansion of the universe is supported by various sets of cosmological observations. Several models proposed in the literature effectively explain the transition from a decelerated phase to the current accelerated phase \cite{RP,M.S.,M.S.-2,T.C.,C.A.,M.C.,A.Y.,T.P.,F2,S10,M2}. In this paper, we present models of the FLRW universe that incorporate a time-varying cosmological term, $\Lambda(t)$, to investigate this transition. Specifically, we have considered a power-law form for the cosmological term as a function of the scale factor: $\Lambda(t)=\Lambda_{0} a(t)^{-\alpha}$, where $\Lambda_{0}$ represents the present value of the cosmological term \cite{Overduin}. We have obtained an exact solution to Einstein's field equations within the framework of $\Lambda(t)$CDM cosmology. Further, we have determined the best-fit values of the model parameters using the combined $H(z)$ + SNe Ia dataset, which are presented in Tab. \ref{tab} and Fig. \ref{Combine}.

In addition to these results, we have conducted a comprehensive analysis of the cosmic expansion of the universe. Our findings reveal that the Hubble function demonstrates a consistently increasing trend with respect to redshift (see Fig. \ref{F_H}). Furthermore, our investigation of the deceleration parameter indicates a smooth transition of $q(z)$ from a phase of decelerated expansion to accelerated expansion at the transition redshift $z_{tr}=0.5^{+0.02}_{-0.02}$ for the best-fit model obtained using the combined $H(z)$ + SNe Ia dataset (see Fig. \ref{F_q}). Therefore, our results align with recent observational data that indicate the present-day acceleration of the universe. From Fig. \ref{F_j}, it is evident that the jerk parameter for the $\Lambda(t)$CDM model remains positive across the entire history of the universe.  Finally, the behaviour of the $Om(z)$ diagnostic, as depicted in Fig. \ref{F_Om}, suggests that our cosmological $\Lambda(t)$CDM model aligns with the phantom scenario. These results indicate that our model is in strong agreement with present-day observations.

\section*{Acknowledgments}
The authors thank the anonymous referees for their helpful comments that improved the
quality of the manuscript.

\section*{Data Availability Statement}
This article does not introduce any new data.

\end{document}